\documentstyle[preprint,aps,prb,epsfig]{revtex}
\begin{document}
\draft
\pagestyle{plain}
\newcommand{\D}{\displaystyle}
\title{\bf  Comment on the anisotropic impurity scattering  
in superconductors}
\author{Grzegorz Hara\'n\cite{AA} and A. D. S. Nagi} 
\vspace{0.4cm}
\address{Department of Physics, 
University of Waterloo,  
Waterloo, Ontario, 
Canada, N2L 3G1}
\date{December 17, 1997}
\maketitle

\begin{abstract}
We discuss a case of a strong anisotropic impurity scattering within the 
model introduced in our previous paper $\left[\right.$Phys. Rev. {\bf B54}, 
15463 (1996)$\left.\right]$ and clarify on our former statement about a possible 
enhancement of the critical temperature in this scattering regime. 
We show, that for the anisotropy of the impurity potential determined 
by the functions from the non identity irreducible representations of the 
crystal point group the critical temperature decreases with the impurity 
scattering rate which is consistent with a generally understood role of disorder 
in superconductors. 
\end{abstract}

\vspace{0.5cm}
\pacs{Keywords: A. high-$T_c$ superconductors, superconductors, D. phase transitions} 

The anisotropy of the impurity potential has become an important issue in the 
interpretation of the critical temperature suppression data in the cuprates. \cite{1,2,3}  
Although still not confirmed by any experiments (a good test may be provided by the angle 
resolved density of states \cite{3a} measurements below $T_c$) the scattering potentials 
assuming a momentum-dependent scattering probability can be very helpful in understanding the     
experimental data. \cite{1,2} In particular, it is possible to obtain within a phenomenological 
model which introduces anisotropy in the impurity scattering potential \cite{1} 
a good agreement with the electron irradiation data in $YBa_2Cu_3O_{7-\delta}$ crystals.  
\cite{4} This model, however, has some interpretational difficulties. 
It has been mentioned \cite{1} that under certain conditions it could lead to 
an enhancement of $T_c$ which seems unphysical  
as superconductivity should be suppressed by an increasing amount of disorder in the 
system. \cite{5} In this letter we address this problem in more detail and show that 
the model really predicts a reduction of the critical temperature if the anisotropy 
of the impurity potential is represented by a certain class of functions.     
This class corresponds to the irreducible non identity representations of the crystal 
point group.  

We discuss a superconducting state defined by the orbital part of the order parameter 
$\Delta\!\left({\bf k}\right)=\Delta e\!\left({\bf k}\right)$ 
where $e\left({\bf k}\right)$ is a real basis function of a one-dimensional (1D)  
irreducible representation of an appropriate point group or a linear combination of 
such functions. For the sake of convenience $e\!\left({\bf k}\right)$ is normalized 
as $\left<e^{2}\right>=1$, 
where $<...>=\int_{FS}dS_k n\left({\bf k}\right)\left(...\right)$ 
denotes the average value over the Fermi surface (FS), 
$n\left({\bf k}\right)$ is the angle resolved FS density 
of states normalized to unity, i.e. $\int_{FS}dS_k n\left({\bf k}\right)=1$,  
and $\int_{FS}dS_{k}$ represents the integration over the Fermi surface.

The impurity scattering potential appropriate for the second-order Born scattering 
limit \cite{5a} is introduced as 

\begin{equation}
\label{e1}
|w\left({\bf k}-{\bf k'}\right)|^{2}=|w_{0}|^{2}+|w_{1}|^{2}
f\left({\bf k}\right)f\left({\bf k'}\right)
\end{equation}

\noindent
with $|w_{0}|$ and $|w_{1}|$ representing isotropic and anisotropic scattering amplitude 
respectively. The momentum-dependent function $f\left({\bf k}\right)$ determines the 
symmetry of the anisotropy of the impurity potential and is assumed to vanish after 
the integration over the Fermi surface, $\left<f\right>=0$. 
Therefore, the Fermi surface average of the scattering potential 
$\left<|w\left({\bf k}-{\bf k'}\right)|^{2}\right>=|w_{0}|^{2}$ is determined by the 
$s$-wave core ($|w_0|$) and the momentum-dependent part in Eq. (\ref{e1}) represents the 
deviations from the isotropic scattering. For the sake of simplicity 
we normalize $f\left({\bf k}\right)$ according to $\left<f^{2}\right>=1$. 
Finally, the requirement of a non negative value of the squared potential in Eq. 
(\ref{e1}) yields $|w_{1}|^{2}\leq |w_{0}|^{2}$ constraint on the potential amplitudes.   

Following standard procedure \cite{1,6,7} the equation for the critical temperature 
of a superconductor with a separable pairing potential 
$V\left({\bf k},{\bf k'}\right)=-V_{0}e\left({\bf k}\right)e\left({\bf k'}\right)$ 
($V_0$ is a positive potential amplitude) 
in the presence of anisotropic impurity scattering is obtained \cite{1}  

\begin{equation}
\label{e2}
\begin{array}{l}
\D\ln\frac{T_c}{T_{c_{0}}}=\left(\left<e\right>^{2}+
\left<ef\right>^{2}-1\right)
\left[\psi\left(\frac{1}{2}+\frac{\Gamma_0}{2\pi T_c}\right)
-\psi\left(\frac{1}{2}\right)\right]+\\
\\
\D\left<ef\right>^{2}\left[\psi\left(\frac{1}{2}\right)-\psi\left(
\frac{1}{2}+\frac{\Gamma_{0}}{2\pi T_{c}}
\left(1-\frac{\Gamma_1}{\Gamma_0}\right)\right)\right]\\
\end{array}
\end{equation}

\noindent
In writing above we have introduced the isotropic $\Gamma_0$  
and anisotropic $\Gamma_1$ impurity scattering rates ($\Gamma_1\leq\Gamma_0$)   

\begin{equation}
\label{e3}
\Gamma_0=\pi N_0n_i|w_0|^2,\;\;\;\;\Gamma_1=\pi N_0n_i|w_1|^2
\end{equation}

\noindent
where $n_{i}$ is impurity (defect) concentration. 
The model introduces two dimensionless parameters which characterize the 
anisotropy of the pair-breaking effect.   
First of them, $\left<ef\right>^2=\left[\int_{FS}dS_{k}n\left({\bf k}\right)
e\left({\bf k}\right)f\left({\bf k}\right)\right]^2$, describes  
the interplay between the pair potential $V\left({\bf k},{\bf k'}\right)$  
and the anisotropic part of the scattering potential,  
$|w\left({\bf k}-{\bf k'}\right)|^{2}$ (Eq. (\ref{e1})). This parameter  
is determined by the symmetry of the superconducting state, 
$e\left({\bf k}\right)$, as well as that of the impurity scattering matrix element,  
$f\left({\bf k}\right)$. According to the normalization of the order 
parameter, $\left<e^2\right>=1$, and the anisotropy function of the impurity 
potential, $\left<f^2\right>=1$, the parameter $\left<ef\right>^2$ takes values 
between 0 and 1. The second parameter in our model, ($\Gamma_1/\Gamma_0$),  
represents the amount of anisotropic scattering rate in impurity 
potential normalized by the isotropic scattering rate (Eq. (\ref{e3})), and its   
value varies also from 0 to 1. 
Several limits of the above equation were examined in Ref. 1. 
Our analysis showed that the symmetry of the anisotropic potential
is an important factor and a significant reduction in the pair-breaking 
strength appears for large values of $\left<ef\right>^2$ with an appropriate level   
of $\Gamma_1/\Gamma_0$. 

Among the terms determining the critical temperature in Eq. (\ref{e2}) only one 
depends on $\Gamma_1/\Gamma_0$. Therefore, $T_c$ can be maximized with respect to 
this parameter by taking 
$\Gamma_1/\Gamma_0$ value maximizing the $\Gamma_1/\Gamma_0$-dependent term. 
This is achieved for $\Gamma_1/\Gamma_0=1$ that is, in the case of the strong  
anisotropic scattering. \cite{1} The critical temperature is then determined by 

\begin{equation}
\label{e4}
\ln\frac{T_c}{T_{c_{0}}}=\left(\left<e\right>^{2}+\left<ef\right>^{2}
-1\right)
\left(\psi\left(\frac{1}{2}+\frac{\Gamma_0}{2\pi T_c}\right)
-\psi\left(\frac{1}{2}\right)\right)
\end{equation}

\noindent
It is easy to see that the critical temperature becomes very robust with respect 
to the impurity scattering rate $\Gamma_0$, especially  
for a significant overlap between $e\left({\bf k}\right)$ and 
$f\left({\bf k}\right)$ functions, that is when $\left<ef\right>^2\sim 1$. 
This case led to the earlier mentioned difficulty of the model. \cite{5} 
The critical temperature is determined now by a coefficient 

\begin{equation}
\label{e4a}
\alpha=\left<e\right>^2+\left<ef\right>^2-1
\end{equation}

\noindent
As long as $\alpha$ is negative 
superconductivity is suppressed by the disorder. However, a positive value of $\alpha$ 
would mean an enhancement of the critical temperature due to impurity scattering. 
Below we show that the second option is impossible for a wide class of functions 
$f\left({\bf k}\right)$ to which we postulate to restrict the model. 

We assume that $f\left({\bf k}\right)$ belongs to a non identity 1D representation   
of the crystal symmetry group or is given by a linear combination of such functions. 
It is worth observing, that the above choice of $f\left({\bf k}\right)$ yields 
$\left<f\right>=0$ which is in agreement with the model. As the next step,  
we express the superconducting order parameter as a sum of a part belonging to the 
identity irreducible representation $e_i\left({\bf k}\right)$ and $e_n\left({\bf k}\right)$   
given by a linear combination of functions transforming according to the non identity 
irreducible representations 
 
\begin{equation}
\label{e5}
e\left({\bf k}\right)=e_i\left({\bf k}\right)+e_n\left({\bf k}\right)
\end{equation}

\noindent
In the case of a $\left(d_{x^2-y^2}+s\right)$-wave superconductor for instance,   
$e_n\left({\bf k}\right)\sim cos2\phi$ and $e_i\left({\bf k}\right)\sim 1$. 
Using the property of orthogonality of different irreducible representations 
of a given symmetry group and taking into account that the FS average defines   
a scalar product in the space of functions defined on the Fermi sheet, we note that  
$\left<e\right>^2=\left<e_i\right>^2$ and $\left<ef\right>^2=\left<e_nf\right>^2$. 
Therefore, the coefficient $\alpha$ can be written as  

\begin{equation}
\label{e6}
\alpha=\left<e_i\right>^2+\left<e_nf\right>^2-1
\end{equation}

\noindent
Through the Schwarz inequality $\left<e_i\right>^2\leq \left<e_i^2\right>$  
we obtain an upper limit on $\alpha$  

\begin{equation}
\label{e7}
\alpha\leq\left<e_i^2\right>+\left<e_nf\right>^2-1
\end{equation}
 
\noindent
It is worth noting, that the equality in the above relation holds only for 
$e_i\left({\bf k}\right)\sim 1$ that is, for instance in the 
$\left(d_{x^2-y^2}+s\right)$-wave or $d_{x^2-y^2}$-wave superconducting state. 
A simple relation $\left<e^2\right>=\left<e_i^2\right>+\left<e_n^2\right>$
and the normalization condition $\left<e^2\right>=1$ yield  
$\left<e_i^2\right>=1-\left<e_n^2\right>$ which leads to an equivalent to the relation 
(\ref{e7}) constraint on $\alpha$  

\begin{equation}
\label{e8}
\alpha\leq\left<e_nf\right>^2-\left<e_n^2\right>
\end{equation}

\noindent
Using a normalization $\left<f^2\right>=1$ we can rewrite the right-hand side 
of the inequality (\ref{e8}) as 

\begin{equation}
\label{e9}
\left<e_nf\right>^2-\left<e_n^2\right>=\left<e_nf\right>^2-
\left<e_n^2\right>\left<f^2\right>
\end{equation}

\noindent
Finally, from the relations (\ref{e8}), (\ref{e9}) and by applying Schwarz inequality 
we get $\alpha\leq 0$. Therefore, according to Eqs. (\ref{e4}) and (\ref{e4a}) $T_c$ is 
a decreasing or a constant function of the impurity scattering rate $\Gamma_0$.  
It is worth observing, that $\left<e_nf\right>^2-\left<e_n^2\right>\left<f^2\right>=0$ 
only when $f\left({\bf k}\right)\sim e_n\left({\bf k}\right)$ i.e. in a 
$d_{x^2-y^2}$- or $\left(d_{x^2-y^2}+s\right)$-wave superconductor for 
$f\left({\bf k}\right)\sim cos2\phi$. Thus, according to the model, these superconducting 
states will not be altered by anisotropic impurity scattering of the 
$d_{x^2-y^2}$-wave symmetry provided the scattering strengths in the isotropic and 
anisotropic channels are equal ($\Gamma_1/\Gamma_0=1$). We believe, that it is rather a   
not plausible situation. 

In conclusion, we have shown that the potential impurity scattering with the 
anisotropy \cite{1} introduced by Eq. (\ref{e1}) leads to a pair-breaking effect  
in superconductors for the anisotropy of the impurity potential,  
$f\left({\bf k}\right)$, determined by a linear combination of functions from the non 
identity irreducible representations of the crystal point group. Therefore, in order 
to reflect a general feature of disorder induced superconductivity suppression, the 
anisotropy of the discussed model potential should be limited to this class of functions. 

This work was supported by the Natural Sciences and Engineering Research Council
of Canada.


\newpage 
\begin{references}
\bibitem[*]{AA} present address: Institute of Physics, Politechnika 
Wroc{\l}awska, Wybrze\.ze Wyspia\'nskiego 27, 50-370 Wroc{\l}aw,  
Poland
\bibitem{1} G. Hara\'n and A. D. S. Nagi, Phys. Rev. {\bf B54}, 15463 (1996)
\bibitem{2} G. Hara\'n and A. D. S. Nagi, preprint 
\bibitem{3} M. L. Kuli\'c and V. Oudovenko, Solid State Commun. {\bf 104}, 375 (1997) 
\bibitem{3a} The effect of impurity induced anisotropy in residual DOS has been suggested  
for another momentum-dependent impurity potential in 
S. Haas, A. V. Balatsky, M. Sigrist, and T. M. Rice, Phys. Rev. 
{\bf B56}, 5108 (1997) 
\bibitem{4} J. Giapintzakis, D. M. Ginsberg, M. A. Kirk, and S. Ockers, 
Phys. Rev. {\bf B 50}, 15967 (1994)
\bibitem{5} G. Hara\'n and A. D. S. Nagi, Bull. Am. Phys. Soc. 42, 715 (1997),  
presentation and a following discussion. 
\bibitem{5a} A more general formulation including multiple scattering processes 
has been given in Ref. 2. 
\bibitem{6} A. A. Abrikosov and L. P. Gorkov, Zh. Eksp. Teor. Fiz. 
{\bf 39}, 1781 (1960) [Sov. Phys. - JETP {\bf 12}, 1243 (1961)]; 
see also   
A. A. Abrikosov, L. P. Gorkov,and I. E. Dzyaloshinski,
{\it Methods of Quantum Field Theory in Statistical Physics}
(Dover, New York, 1975), sec 39
\bibitem{7} K. Maki, in {\it Superconductivity}, R. D. Parks (ed.), 
(Marcel Dekker, New York, 1969), Vol. 2, pp. 1035-1102
\end{references}
\end{document}